\documentclass[showpacs,preprintnumbers,twocolumn,amsmath,amssymb]{revtex4}
\usepackage{graphicx}
\usepackage{dcolumn}
\usepackage{bm}
\usepackage{epsfig}

\newcommand{\dhd}{{\textstyle d}
\lower.03ex\hbox{\kern-0.40em$^{\scriptstyle-}$}\kern-0.08em{}}  

\newcommand{\half}{{1\over 2}}
\newcommand{\bu}{{\bullet}}

\newcommand{\barz}{{\bar z}}

\newcommand{\calo}{{\cal O}}

\begin{document}

\preprint{JLAB-THY-08-923}

\title{
Conformal kernel for NLO BFKL equation in ${\cal N}$=4 SYM}

\author{Ian Balitsky and Giovanni A. Chirilli}
\affiliation{
Physics Dept., ODU, Norfolk VA 23529, \\
and \\
Theory Group, Jlab, 12000 Jefferson Ave, Newport News, VA 23606
}
\email{balitsky@jlab.org, chirilli@jlab.org}

\begin{abstract}
Using the requirement of M\"{o}bius invariance of ${\cal N}$=4 SYM amplitudes in the Regge limit 
we restore the conformal NLO BFKL kernel out of the eigenvalues known from the forward NLO BFKL
result.
\end{abstract}


\maketitle


The high-energy behavior of perturbative amplitudes is given by the BFKL pomeron \cite{bfkl}. In the leading order, the BFKL
equation is conformally invariant under the M\"{o}bius SL(2,C) group of transformations  of the transverse plane.
In  the next-to-leading order (NLO) the BFKL kernel in QCD is not invariant because of the running coupling, but the kernel in 
${\cal N}=4$ SYM is expected to be invariant. The eigenvalues of this conformal kernel are known from the 
calculation of forward NLO BFKL in the momentum space \cite{lipkot}. In a conformal theory it is possible to recover the 
amplitude of the non-forward scattering of two reggeized gluons from the forward scattering amplitude.  Using the NLO kernel
for evolution of color dipoles in QCD \cite{nlobk} we guess the M\"{o}bius invariant kernel for ${\cal N}$=4 SYM and check that it reproduces
known eigenvalues \cite{lipkot}.

At high energies the typical forward scattering amplitude has the form
\begin{eqnarray}
&&\hspace{-3mm}
A(s,0)~=~s\!\int\!{d^2 q\over q^2}
{d^2 q'\over {q'}^2}F_A(q)F_B(q')
\label{lip1}\\
&&\hspace{-3mm} \times
\!\int_{a-i\infty}^{a+i\infty}\!{d\omega\over 2\pi i}~f_+(\omega)\Big({s\over qq'}\Big)^\omega
G_\omega(q,q')
\nonumber
\end{eqnarray}
where  $F_A(q)$, $F_B(q')$ are the impact factors, $f_+(\omega)={e^{i\pi\omega}-1\over\sin\pi\omega}$ is the signature factor, and $G_\omega(q,q')$ is the partial wave of the forward reggeized gluon scattering amplitude satisfying the 
BFKL equation
\begin{equation}
\omega G_\omega(q,q')=\delta^{2}(q-q')+\int\! d^2p K(q,p)G_\omega(p,q') 
\label{wgw}
\end{equation}
In ${\cal N}=4$ SYM the kernel $K(q,p)$ is known up to the 
next-to-leading order \cite{lipkot} 
\begin{eqnarray}
&&\hspace{-3mm}
\int\! d^2p K(q,p)f(p)
\label{nlobfkernelmom}\\
&&\hspace{-3mm} =~
{\alpha_sN_c\over\pi^2}
\!\int\! d^2p~
\Big\{{1\over (q-p)^2}\Big(1-{\alpha_sN_c\pi\over 12}\Big)
[f(p)-{q^2\over 2p^2}f(q)]
\nonumber\\
&&\hspace{-3mm} 
+~{\alpha_sN_c\over 4\pi}
\big[\Phi(q,p)-{\ln^2{q^2/p^2}\over (q-p)^2}\big]~f(p)\Big\} 
+{3\alpha_s^2N_c^2\over 2\pi^2}\zeta(3)f(p)
\nonumber
\end{eqnarray}
where $\zeta$ is the Riemann zeta-function and
\begin{eqnarray}
&&\hspace{-4mm}
\Phi(q,p)~=~
{(q^2-p^2)\over (q-p)^2(q+p)^2}
\Big[\ln{q^2\over p^2}\ln{q^2p^2(q-p)^4\over (q^2+p^2)^4}
\nonumber\\
&&\hspace{-4mm}
+~2{\rm Li}_2\big(-{p^2\over q^2}\big)-2{\rm Li}_2\big(-{q^2\over p^2}\big)\Big]
-\Big[1-{(q^2-p^2)^2\over (q-p)^2(q+p)^2}\Big]
\nonumber\\
&&\hspace{-4mm}
\times~\Big[\!\int_0^1-\int_1^\infty\Big]
{du\over (qu-p)^2}\ln{u^2q^2\over p^2}
\label{dephi}
\end{eqnarray}
Here Li$_2$ is the dilogarithm.

The eigenvalues of the kernel (\ref{nlobfkernelmom}) are \cite{lipkot}
\begin{eqnarray}
&&\hspace{-2mm}
\int\! d^2p \Big({p^2\over q^2}\Big)^{-\half+i\nu}e^{in\phi}K(q,p)~=~\omega(n,\nu),~~~~
\label{eigen1}\\
&&\hspace{-2mm} 
\omega(n,\nu)~=~{\alpha_s\over \pi}N_c\Big[\chi(n,\half+i\nu)+{\alpha_sN_c\over 4\pi}\delta(n,\half+i\nu)
\Big],
\nonumber\\
&&\hspace{-1mm}
\delta(n,\gamma)~=~
6\zeta(3)-
{\pi^2\over 3}\chi(n,\gamma)-\chi"(n,\gamma)
\nonumber\\
&&\hspace{22mm}
-~2\Phi(n,\gamma)-2\Phi(n,1-\gamma)
\nonumber
\end{eqnarray}
where $\chi(n,\gamma)=2\psi(1)-\psi(\gamma+{n\over 2})-\psi(1-\gamma+{n\over 2})$ and
\begin{eqnarray}
&&\hspace{-1mm}
\Phi(n,\gamma)~=~\int_0^1\!{dt\over 1+t}~t^{\gamma-1+{n\over 2}}
\Big\{{\pi^2\over 12}-{1\over 2}\psi'\Big({n+1\over 2}\Big)
\nonumber\\
&&\hspace{-2mm}-{\rm Li}_2(t)-{\rm Li}_2(-t)
-~\Big(\psi(n+1)-\psi(1)+\ln(1+t)
\nonumber\\
&&\hspace{-1mm}
+~\sum_{k=1}^\infty{(-t)^k\over k+n}\Big)\ln t-\sum_{k=1}^\infty{t^k\over (k+n)^2}[1-(-1)^k]\Big\}
\label{fi}
\end{eqnarray}

The Regge limit of the amplitude $A(x,y;x',y')$ in the coordinate space can be achieved as 
\begin{eqnarray}
&&\hspace{-1mm}
x=\lambda x_\ast p_1+x_\perp,~~~ y=\lambda y_\ast p_2+y_\perp, ~~~~~~~
\nonumber\\
&&\hspace{-1mm}
x'=\rho x_\bu p_2 +x'_\perp,~~~  y'=\rho y'_\bu p_2 +y'_\perp
\label{reggelimit}
\end{eqnarray}

 with $\lambda,\rho\rightarrow\infty$ and $x_\ast>0>y_\ast$, $x'_\bu>0>y'_\bu$. Hereafter we use the notations $x_\bu=p_1^\mu x_\mu$, $x_\ast=p_2^\mu x_\mu$
where $p_1$ and $p_2$  are light-like vectors normalized by $2(p_1,p_2)=s$. These ``Sudakov variables'' are related to 
the usual light-cone coordinates
$x^\pm={1\over\sqrt{2}}(x^0\pm x^3)$ by $x_\ast=x^+\sqrt{s/2},~x_\bu=x^-\sqrt{s/2}$ so $x={2\over s}x_\ast p_1+{2\over s}x_\bu p_2+x_\perp$. We use the (1,-1,-1,-1) metric  so $x^2={4\over s}x_\bu x_\ast -\vec{x}_\perp^2$.

In the Regge limit (\ref{reggelimit}) the full conformal group  reduces to M\"{o}bius subgroup SL(2,C) leaving the transverse plane $(0,0,z_\perp)$ invariant. 
In a conformal theory the four-point amplitude $A(x,y;x',y')$ depends on two conformal ratios which can be chosen as
\begin{eqnarray}
&&\hspace{-1mm}
R~=~{(x-x')(y-y')^2\over (x-y)^2(x'-y')^2},~~~~
\nonumber\\
&&\hspace{-1mm}r~=~R\Big[1-{(x-y')^2(y-x')^2\over (x-x')^2(y-y')^2}+{1\over R}\Big]^2
\label{cratios1}
\end{eqnarray}
The conformal ratio $R$ scales as $\lambda^2\rho^2$ while $r$ is does not depend on $\lambda$ or $\rho$.  
Following Ref. \cite{penecostalba} (see also Ref. \cite{penedones}) it is convenient to introduce two SL(2,C)-invariant vectors
\begin{eqnarray}
&&\hspace{-5mm}
\kappa~=~{\sqrt{s}\over 2x_\ast}(p_1-{x^2\over s}p_2+x_\perp)-{\sqrt{s}\over 2y_\ast}(p_1-{y^2\over s}p_2+y_\perp)
\label{kappas}\\
&&\hspace{-5mm}
\kappa'~=~{\sqrt{s}\over 2x'_\bu}(p_1-{{x'}^2\over s}p_2+x'_\perp)-{\sqrt{s}\over 2y'_\bu}(p_1-{{y'}^2\over s}p_2+y'_\perp)
\nonumber
\end{eqnarray}
such that 
\begin{equation}
\kappa^2{\kappa'}^2~=~{1\over R}~~~~~~{\rm and}~~~~~~~~4(\kappa\cdot\kappa')^2~=~{r\over R}
\label{Rr}
\end{equation}
(here $x^2=-x_\perp^2,~{x'}^2=-{x'}_\perp^2$ and similarly for $y$). In the coordinate space the analog of Eq. (\ref{lip1})  has the form: 
\begin{eqnarray}
&&\hspace{-7mm}
A(x,y;x',y')~=~\int\! d^2z_1 d^2z_2d^2z'_1 d^2z'_2~I_A(x,y;z_1,z_2)
\nonumber\\
&&\hspace{-7mm}
\times~\!\int\! {d\omega\over 2\pi} ~R^{\omega\over 2} f_+(\omega)G_\omega(z_1,z_2;z'_1,z'_2) I_B(x',y';z'_1,z'_2)
\label{coord1}
\end{eqnarray}
where the partial wave of the reggeized gluon scattering amplitude satisfies the equation
\begin{eqnarray}
&&\hspace{-5mm}
\omega G_\omega(z_1,z_2;z'_1,z'_2)~=~\ln^2 {(z_1-z'_1)^2(z_2-z'_2)^2\over (z_2-z'_1)^2(z_1-z'_2)^2}
\nonumber\\
&&\hspace{-5mm}
+~
\int\! d^2t_1d^2t_2 ~K(z_1,z_2;t_1,t_2)G_\omega(t_1,t_2;z'_1,z'_2)
\label{wgwcoord}
 \end{eqnarray}
Here the  first term in the r.h.s. is the leading-order contribution coming from two-gluon exchange.

The meaning of the Eq. (\ref{coord1}) is that the amplitude is factorized into the product of 
three terms $I_A$, $I_A$, and $G_\omega$ corresponding to rapidities $\eta\sim\eta_A$, 
$\eta\sim\eta_B$, and $\eta_A>\eta>\eta_B$, respectively. With conformally invariant factorization  
of the amplitude into such product
the impact factors and $G_\omega$ should be separately M\"{o}bius invariant leading
 to invariant kernel $K(z_1,z_2;t_1,t_2)$.   The eigenfunctions of a conformal kernel are \cite{lip86}
\begin{equation}
\hspace{-0mm}
E_{\nu,n}(z_{10},z_{20})~
=~\Big[{\tilde{z}_{12}\over \tilde{z}_{10}\tilde{z}_{20}}\Big]^{\half+i\nu+{n\over 2}}
\Big[{\barz_{12}\over \barz_{10}\barz_{20}}\Big]^{\half+i\nu-{n\over 2}}
\label{eigenfunctions}
\end{equation}
where $\tilde{z}=z_x+iz_y,\barz=z_x-iz_y$ and $z_{10}\equiv z_1-z_0$ etc. Denoting the eigenvalues of the kernel $K$ by $\omega(n,\nu)$ 
\begin{eqnarray}
&&\hspace{-3mm}
\int\! d^2t_1d^2t_2 ~K(z_1,z_2;t_1,t_2)E_{\nu,n}(t_1-z_0,t_2-z_0)~
\nonumber\\
&&\hspace{-1mm}
=~\omega(n,\nu)E_{\nu,n}(z_{10},z_{20})
\label{eigenvalues}
\end{eqnarray}
and substituting the formal solution of the Eq. (\ref{wgwcoord}) into Eq. (\ref{coord1}) we obtain 
\begin{eqnarray}
&&\hspace{-4mm}
A(x,y;x',y')~=\sum_{n=-\infty}^\infty\!\int\! {d\nu\over\pi^2}~ 
{-2\big(\nu^2+{n^2\over 4}\big)R^{\half\omega(n,\nu)}\over [\nu^2+{(n-1)^2\over 4}] [\nu^2+{(n+1)^2\over 4}]}
\nonumber\\
&&\hspace{-4mm}
\times~f_+(\omega(n,\nu))\!
\!\int\! d^2z_0 d^2z_1 d^2z_2 I_A(x,y;z_1,z_2)E_{\nu,n}(z_{10},z_{20})
\nonumber\\
&&\hspace{-4mm} 
\times~\!\int\! d^2z'_1 d^2z'_2~I_B(x',y';z'_1,z'_2)E^\ast_{\nu,n}(z'_1-z_0,z'_2-z_0) 
\nonumber
\end{eqnarray}
As demonstrated in Ref. \cite{penecostalba} an impact factor depends  on one conformal (M\"{o}bius invariant) ratio
\begin{eqnarray}
&&\hspace{-1mm}
I_A(x,y;z_1,z_2)={1\over z_{12}^4}
I_A\Big({\kappa^2(\zeta_1\cdot \zeta_2)\over 2(\kappa\cdot \zeta_1)(\kappa\cdot \zeta_2)}\Big), 
~~~~~~
\nonumber\\
&&\hspace{-1mm}
I_B(x',y';z'_1,z'_2)={1\over {z'}_{12}^4} I_B\Big({{\kappa'}^2(\zeta'_1\cdot \zeta'_2)\over  2(\kappa'\cdot \zeta'_1)(\kappa'\cdot \zeta'_2)}\Big)
\nonumber
\end{eqnarray}
where $\zeta_1\equiv p_1+{z_{1\perp}^2\over s}p_2+z_{1\perp}$ and similarly for other $\zeta$'s. This enables us to carry out the integrations over $z_i$ and $z'_i$. 
The formulas are especially simple when we consider the correlator of four scalar currents such as ${\rm Tr\{Z^2\}}$ ($Z={1\over\sqrt{2}}(\phi_1+i\phi_2)$) 
so that only the term with $n=0$ contributes. From conformal (M\"{o}bius) invariance we get \cite{penecostalba} 
\begin{eqnarray}
&&\hspace{-1mm}
\int\! {d^2z_1d^2z_2\over z_{12}^4} ~ I_A\Big({\kappa^2(\zeta_1\cdot \zeta_2)\over 2(\kappa\cdot \zeta_1)(\kappa\cdot \zeta_2)}\Big)
\Big({z_{12}^2\over z_{10}^2z_{20}^2}\Big)^{\half +i\nu}~
\nonumber\\
&&\hspace{-1mm}=~{1+4\nu^2\over 8\pi}
{\Gamma^2\big(\half-i\nu\big)\over\Gamma(1-2i\nu)}
\Big({\kappa^2\over 4(\kappa\cdot\zeta_0)^2}\Big)^{\half +i\nu}I_A(\nu)
\label{impactor1}
\end{eqnarray}
 (here $\zeta_0\equiv p_1+{z_{0\perp}^2\over s}p_2+z_{0\perp}$) and therefore (cf. Ref. \cite{penecostalba})
\begin{eqnarray}
&&\hspace{-3mm}
(x-y)^4(x'-y')^4\langle \calo(x) \calo^\dagger(y)\calo(x')\calo^\dagger(y') \rangle~
\label{koppinkoop}\\
&&\hspace{-3mm}
=~-{1\over 2}\!\int\! d\nu~ f_+(\nu)
{\tanh\pi\nu\over \nu}I_A(\nu)I_B(-\nu)
\Omega(r,\nu)R^{\half\omega(\nu)}
\nonumber
\end{eqnarray}
where ${\calo}\equiv{4\pi^2\sqrt{2}\over \sqrt{N_c^2-1}}{\rm Tr\{Z^2\}}$,  $\omega(\nu)\equiv \omega(0,\nu)$,  $f_+(\nu)\equiv f_+(\omega(\nu))$,
and 
\begin{eqnarray}
&&\hspace{-5mm}
\Omega(r,\nu)~=~{\nu^2\over\pi^3}
\!\int\! d^2z \Big({\kappa^2\over (\kappa\cdot\zeta)^2}\Big)^{\half +i\nu} \Big({{\kappa'}^2\over (\kappa'\cdot\zeta)^2}\Big)^{\half -i\nu}
\label{integral7}
\end{eqnarray}
(Since the integral (\ref{integral7}) does not scale with $\lambda,\rho$ it can depend only on 
${(\kappa\cdot\kappa')^2\over \kappa^2{\kappa'}^2}={r\over 4}$). 
The equation (\ref{koppinkoop}), obtained in Ref. \cite{cornalba} from general consideration of the Regge limit in a conformal theory, 
proves the existence ot the conformally invariant factorization (\ref{coord1}).
Note that all the dependence on large energy ($\equiv$ large $\lambda,\rho$) is contained in $R^{\half\omega(\nu)}$. 
For completeness, let us mention that in the leading order in perturbation theory 
$I(\nu)={2\pi^2\alpha_s\over\cosh\pi\nu} \sqrt{1-{1\over N_c^2}}$.

To restore the NLO BFKL 
kernel in the coordinate representation (\ref{coord1}) from the eigenvalues (\ref{eigen1}) in the momentum representation we must prove that 
Eq. (\ref{koppinkoop}) agrees with Eq. (\ref{lip1}) with the same set of $\omega(\nu)$. (Strictly speaking, we need to demonstrate this property 
for arbitrary $n$ but here we will do it only for $n=0$).

 In order to perform Fourier transformation of the correlator (\ref{koppinkoop}) we need to relax the 
 limit (\ref{reggelimit}) by allowing small $x_\bu\sim y_\bu\sim 1/\lambda$
 and $x'_\ast\sim y'_\ast\sim 1/\rho$. 
 The conditions  (\ref{Rr})  for vectors (\ref{kappas}) are now satisfied  up to ${1\over \lambda^2}$ and ${1\over \rho^2}$ corrections. 
 The correlator  (\ref{koppinkoop})  takes the form
\begin{eqnarray}
&&\hspace{-4mm}
 (x-y)^4(x'-y')^4\langle \calo(x)\calo^\dagger(y)\calo(x')\calo^\dagger(y')\rangle~
 \label{koppelatop}\\
&&\hspace{-4mm}=~
{-1\over 2\pi^3}\!\int\! d\nu ~f_+(\nu) \nu\tanh \pi\nu
\Big[{16x_\ast y_\ast x'_\bu y'_\bu\over s^2(x-y)^2(x'-y')^2}\Big]^{\half\omega(\nu)}
\nonumber\\
&&\hspace{-4mm}
\times~
\!\int\! d^2z_0~\Bigg[{{(x-y)^2\over x_\ast y_\ast}\over \big(
{(x-z_0)_\perp^2\over x_\ast} 
-{(y-z_0)_\perp^2\over y_\ast}-{4\over s}(x-y)_\bu\big)^2}\Bigg]^{\half +i\nu}\!\!\!I_A(\nu)
\nonumber\\
&&\hspace{-4mm}
\times~\Bigg[{-{(x'-y')^2\over x'_\bu|y'_\bu|}\over \big(
{(x'-z_0)_\perp^2\over x_\bu} 
-{(y'-z_0)_\perp^2\over y'_\bu}-{4\over s}(x'-y')_\ast\big)^2}\Bigg]^{\half -i\nu}\!\!\!I_B(-\nu)
\nonumber
\end{eqnarray}
The forward scattering amplitude can be defined as (cf. Ref. \cite{mobzor})
\begin{eqnarray}
&&\hspace{-4mm}
A(s,0)~
\nonumber\\
&&\hspace{-4mm}
=~-i\!\int\! 
\!d^4zd^4x d^4y
\langle \calo(x_\bu,x_\ast +z_\ast,x_\perp+z_\perp)
\calo^\dagger(0,z_\ast,z_\perp)
\nonumber\\
&&\hspace{-4mm}
\times~\calo(y_\bu+z_\bu,y_\ast,y_\perp)\calo^\dagger(z_\bu,0,0)\rangle ~e^{-ip_A\cdot x-ip_B\cdot y}
\label{forwardamplitude}
\end{eqnarray}
where  $p_A=p_1+{p_A^2\over s}p_2$ and $p_B=p_2+{p_B^2\over s}p_1$. Substituing Eq. (\ref{koppelatop}) in Eq. (\ref{forwardamplitude}) 
and performing the integrations over  the coordinates we obtain
\begin{eqnarray}
&&\hspace{-3mm}
A(s,0)=~{8\pi^2s\over (p_A^2p_B^2)^{3/2}}\!\int\! d\nu ~I_A(\nu)I_B(-\nu)
\Big({s\over 4 \sqrt{p^2_Ap^2_B}}\Big)^{\omega(\nu)}
\nonumber\\
&&\hspace{-3mm}
\times~f_+(\nu)\Big({ p_A^2\over p_B^2}\Big)^{i\nu}
\left|
{\Gamma^2\big({3\over 2}+{\omega(\nu)\over 2}+i\nu\big)
\over \Gamma(3+\omega(\nu)+2i\nu)}{\Gamma^2\big(\half-i\nu\big)\over \Gamma(1-2i\nu)}\right|^2
\label{adamom}
\end{eqnarray}
This should be compared with Eq. (\ref{lip1}) which takes the form
\begin{eqnarray}
&&\hspace{-4mm}
A(s,0)~=~s\int\! {d^2q\over q^2}{d^2q'\over {q'}^2}F_A(q)F_B(q')
\label{lip2}\\
&&\hspace{-4mm}
\times~
\int\!{d\nu\over 2\pi^2}~f_+(\nu)(q^2)^{-\half+{i\nu\over 2}}({q'}^2)^{-\half-{i\nu\over 2}}\Big({s\over |q||q'|}\Big)^{\omega(\nu)}
\nonumber\\
&&\hspace{-4mm}
=~\!\int\!{d\nu\over 2\pi^2}{sf_+(\nu)\over (p_A^2p_B^2)^{3\over 2}}F_A(\nu)F_B(-\nu)\Big[{s\over 4 \sqrt{p^2_Ap^2_B}}\Big]^{\omega(\nu)}\Big({ p_A^2\over p_B^2}\Big)^{i\nu}
\nonumber
\end{eqnarray}
It is clear that Eq. (\ref{adamom}) and Eq. (\ref{lip2}) coincide after the redefinition of the impact factor
\begin{equation}
\hspace{-4mm}
F_A(\nu)~
\nonumber\\
\hspace{-1mm}=~4\pi^2I_A(\nu)
{\Gamma^2\big({3\over 2}-i\nu+{\omega(\nu)\over 2}\big)
\Gamma^2\big({1\over 2}+i\nu\big)
\over \Gamma(3-2i\nu+\omega(\nu)) \Gamma(1+2i\nu)}
\nonumber
\end{equation}
and similarly for $F_B$.

Now we are in a position to restore $K_{\rm NLO}(z_1,z_2;t_1,t_2)$ from the eigenvalues (\ref{eigen1}).
At the leading-order level $K$ is given by the BFKL kernel in the dipole form (the linear part of the BK equation \cite{npb96,yura})
\begin{eqnarray}
&&\hspace{-4mm}
K_{\rm LO}(z_1,z_2;z_3,z_4)~=~{\alpha_sN_c\over 2\pi^2}\Big[{z_{12}^2\delta^{2}(z_{13})\over z_{14}^2z_{24}^2}
+{z_{12}^2\delta^{2}(z_{24})\over z_{13}^2z_{23}^2}
\nonumber\\
&&\hspace{-4mm}
-~\delta^{2}(z_{13})\delta^{2}(z_{24})\!\int\! d^2z~{z_{12}^2\over (z_1-z)^2(z_2-z)^2}\Big]
\label{klo}
\end{eqnarray}
Using the eigenvalues $\omega(n,\nu)$ and the requirement of conformal invariance it is possible to restore the conformal kernel 
for the NLO BFKL equation
\begin{eqnarray}
&&\hspace{-3mm}
K_{\rm NLO}(z_1,z_2;z_3,z_4)~=~-{\alpha_sN_c\over 4\pi}{\pi^2\over 3}K_{\rm LO}(z_1,z_2;z_3,z_4)
\nonumber\\
&&\hspace{-3mm}
+~{\alpha_s^2N_c^2\over 8\pi^4}~\Big\{ {z_{12}^2\over z_{34}^2z_{13}^2z_{24}^2}
\Big[\Big(
1+{z_{12}^2z_{34}^2\over z_{13}^2z_{24}^2- z_{14}^2z_{23}^2}\Big)\ln{z_{13}^2z_{24}^2\over z_{14}^2z_{23}^2}\nonumber\\
&&\hspace{-3mm} +~
2\ln{z_{12}^2z_{34}^2\over z_{14}^2z_{23}^2}\Big]+12\pi^2\zeta(3)\delta(z_{13})\delta(z_{24})
\Big\}
\label{nlokonf}
\end{eqnarray}
Eq. (\ref{nlokonf}) is the main result of the present paper. It is worth noting that the first term in braces in the l.h.s. corresponds
to the analytic term in the conformal part of NLO BK kernel in QCD \cite{nlobk}.

 The equation (\ref{wgwcoord}) with the kernel (\ref{nlokonf}) is obviously conformally invariant. Let us prove that its eigenvalues are given by Eq. (\ref{eigen1}). The integral
\begin{eqnarray}
&&\hspace{-3mm}
\int\! d^2z_3 d^2z_4 K_{\rm NLO}(z_1,z_2;z_3,z_4)E_{n,\nu}(z_{30},z_{40})~
\label{zeintegral}\\
&&\hspace{-3mm}=~[c(n,\nu)+{\alpha_s^2N_c^2\over 4\pi^2}\big(6\zeta(3)-{\pi^2\over 3}\chi(n,\nu)\big)]E_{n,\nu}(z_{10},z_{20})
\nonumber
\end{eqnarray}
can be reduced to
\begin{eqnarray}
&&\hspace{-4mm} 
{\alpha_s^2N_c^2\over 8\pi^4}
\!\int \!dz_3d z_4{z_{12}^2\over z_{34}^2z_{13}^2z_{24}^2}
\Big\{
2\ln{z_{12}^2z_{34}^2\over z_{14}^2z_{23}^2}+
\Big[
{z_{12}^2z_{34}^2\over z_{13}^2z_{24}^2- z_{14}^2z_{23}^2}
\nonumber\\
&&\hspace{-4mm} 
+~1\Big]\ln{z_{13}^2z_{24}^2\over z_{14}^2z_{23}^2}\Big\}
~\big({\tilde{z}_{34}\over \tilde{z}_{12}}\big)^{\half+i\nu-{n\over 2}}\big({\barz_{34}\over \barz_{12}}\big)^{\half+i\nu+{n\over 2}}
~=~c(n,\nu)
\nonumber
\end{eqnarray}
by setting $z_0=0$ and making the inversion $x_i\rightarrow x_i/x^2$. Taking now $z_2=0$ we obtain
\begin{eqnarray}
&&\hspace{-2mm}                              
{\alpha_s^2N_c^2\over 8\pi^4}
\!\int \!d^2 z~{z_1^2\over z^2}
\big({\tilde{z}\over \tilde{z}_1}\big)^{\half+i\nu-{n\over 2}}\big({\barz\over\barz_1}\big)^{\half+i\nu+{n\over 2}} 
\int\! d^2z' 
\nonumber\\
&&\hspace{-3mm}{1\over(z_1-z-z')^2{z'}^2}~\Big\{
2\ln{z_1^2z^2\over (z_1-z')^2(z+z')^2}
\nonumber\\
&&\hspace{-2mm} 
+~\Big[1+{z_1^2z^2\over (z_1-z-z')^2{z'}^2- (z_1-z')^2(z+z')^2}\Big]
\nonumber\\
&&\hspace{-2mm} 
\times~\ln{(z_1-z-z')^2{z'}^2\over (z_1-z')^2(z+z')^2}\Big\}                    
~=~c(n,\nu)
\label{eigenvalues2}
\end{eqnarray}
Using now the integral
\begin{equation}
\!\int\! {d^2z'\over \pi}~{\ln{(z_1-z')^2(z+z')^2/(z_1^2z^2)}\over(z_1-z-z')^2{z'}^2}~=~{1\over (z_1-z)^2}\ln^2{z^2\over z_1^2}
\nonumber
\end{equation}
and the integral $J_{13}$ from Ref. \cite{integral}
\begin{eqnarray}
&&\hspace{-7mm}
\int\! {d^2z'\over 2\pi} ~
\Big[1+{z_1^2z^2\over (z_1-z-z')^2{z'}^2-(z_1-z')^2(z+z')^2}\Big]
\nonumber\\
&&\hspace{-7mm}
\times~{{z'}^{-2}\over (z_1-z-z')^2}\ln{(z_1-z-z')^2{z'}^2\over (z_1-z')^2(z+z')^2}~
=~\Phi(z_1,z)
\nonumber
\end{eqnarray}
(see Eq. (\ref{dephi}) for the definition of $\Phi$) we obtain
\begin{eqnarray}
&&\hspace{-2mm}                              
{\alpha_s^2N_c^2\over 4\pi^3}
\!\int \!d^2 z~ (z^2/z_1^2)^{-\half+i\nu+{n\over 2}}e^{-in\phi}
\Big[ -{1\over (z_1-z)^2}\ln^2{z^2\over z_1^2}          
\nonumber\\
&&\hspace{-1mm}+~\Phi(z_1,z)\Big] ~=~c(n,\nu)
\label{eigenvalues3}
\end{eqnarray}
where $\phi$ is the angle  between $\vec{z}$ and $\vec{z}_1$.
The final step is to use integrals \cite{lipkot}
\begin{eqnarray}
&&\hspace{-3mm}                              
\!\int \!{d^2 z\over\pi}~{1\over (z_1-z)^2}(z^2/z_1^2)^\gamma e^{in\phi}
\ln^2{z^2\over z_1^2}~=~\chi''(n,\gamma)
\nonumber\\
&&\hspace{-3mm}   
\int\! {d^2z\over 2\pi}~\big({z^2\over z_1^2}\big)^{\gamma-1}e^{in\phi}
\Phi(z_1,z)~=~-\Phi(n,\gamma)-\Phi(n,1-\gamma)
\nonumber
\end{eqnarray}
 Comparing to 
Eq. (\ref{nlobfkernelmom}) we see that $c(n,\nu)={\alpha_s^2N_c^2\over 4\pi^2}\big[-\chi''(n,\nu)-2\Phi(n,\half+i\nu)-2\Phi(\half-i\nu)\big]$ which corresponds to 
$\omega_{\rm NLO}$ from Eq. (\ref{eigen1})

Let us comment on the result  in the literature that NLO BFKL in the coordinate space is not conformally invariant \cite{fadinlo}. 
As we mentioned above, the conformal result for the NLO BFKL kernel (\ref{nlokonf}) corresponds to the factorization in rapidity
consistent with M\"{o}bius  invariance. In other words, this kernel should describe the evolution of the color dipole with the 
conformally invariant rapidity cutoff. At present, there is no obvious way to impose such a cutoff although we believe that it can
be done by constructing a ``composite operator''  for a color dipole, order by order in the perturbation theory. We also think
that the Fourier transform of Eq. (\ref{koppelatop}) in the non-forward case whould give the precise cutoff for the longitudinal integrations 
in the momentum space and cure the discrepancy with the results of Ref. \cite{fadinlo}.

One can also restore the NLO QCD kernel with the same  rapidity cutoff  implicitly defined above to satisfy the requirement of  the conformal invariance 
of the ${\cal N}=4$ kernel (\ref{nlokonf}). Using the results of \cite{prd75,nlobk} 
 one obtains
\begin{eqnarray}
&&\hspace{-4mm}
K^{\rm QCD}_{\rm NLO}(z_1,z_2;z_3,z_4)~=~K_{\rm NLO}(z_1,z_2;z_3,z_4)
\label{nloqcd}\\
&&\hspace{-4mm}
+~{\alpha_s\over 4\pi}\big(b\ln z_{12}^2\mu^2+{67\over 9}N_c-{10\over 9}n_f\big)K_{\rm LO}(z_1,z_2;z_3,z_4)
\nonumber\\
&&\hspace{-4mm}
+~{\alpha_s^2N_c\over 8\pi^3}b\Big[\delta^{2}(z_{13})\big({1\over z_{14}^2}-{1\over z_{24}^2}\big)\ln{z_{14}^2\over z_{24}^2}
+\delta^{2}(z_{24})\big({1\over z_{13}^2}
\nonumber\\
&&\hspace{-4mm}
-~{1\over z_{23}^2}\big)\ln{z_{13}^2\over z_{23}^2}
-\delta^{2}(z_{13})\delta^{2}(z_{24})\!\int\! d^2z_0\big({1\over z_{10}^2}-{1\over z_{20}^2}\big)\ln{z_{10}^2\over z_{20}^2}\Big]
\nonumber\\
&&\hspace{-4mm}
+~{\alpha_s^2N_c^2\over 8\pi^4z_{34}^4}\Bigg[
-~3{z_{12}^2z_{34}^2\over z_{13}^2z_{24}^2-z_{14}^2z_{23}^2}
\ln{z_{13}^2z_{24}^2\over z_{14}^2z_{23}^2}
\nonumber\\
&&\hspace{-4mm}
+~\big(1+{n_f\over N_c^3}\big)\Big({z_{13}^2z_{24}^2+z_{14}^2z_{23}^2-z_{12}^2z_{34}^2\over z_{13}^2z_{24}^2-z_{14}^2z_{23}^2}
\ln{z_{13}^2z_{24}^2\over z_{14}^2z_{23}^2}-2\Big)
\Bigg]
\nonumber
\end{eqnarray}
where $b=11N_c/3-2n_f/3$ and $\mu$ is the normalization point in the $\overline{\rm MS}$ scheme. This kernel has the QCD eigenvalues $\omega(n,\nu)$ from Ref. \cite{nlobfkl}. Note that Eq. (\ref{nloqcd}) is different from the 
NLO BK kernel for the evolution of color dipoles in Ref. \cite{nlobk} since the ``rigid cutoff'' $\alpha>\sigma$ adopted in that paper is not conformally invariant.

The authors are grateful to L.N. Lipatov and J. Penedones for valuable discussions. 
This work was supported by contract
 DE-AC05-06OR23177 under which the Jefferson Science Associates, LLC operate the Thomas Jefferson National Accelerator Facility.
 G.A.C. work was supported by DOE grant DE-FG02-97ER41028 and by the
Jefferson Lab Graduate Fellowship.

 

\section*{References}

\vspace{-5mm}
\begin{thebibliography}{99}

\bibitem{bfkl}
V.S. Fadin, E.A. Kuraev, and L.N. Lipatov,
{\it Phys. Lett.} {\bf B 60}, 50 (1975);
I. Balitsky and L.N. Lipatov,
{\it Sov. Journ. Nucl. Phys.} 
{\bf 28}, 822 (1978).


\bibitem{lipkot}
 A.V. Kotikov and L.N. Lipatov, 
{\it Nucl. Phys.}  {\bf B582}, 19 (2000);
{\it Nucl. Phys.}  {\bf B5661}, 19 (2003).
 Erratum-ibid., {\bf B685}, 405 (2004).
 
\bibitem{nlobk}
I. Balitsky and G.A. Chirilli,
{\it  Phys.Rev.} {\bf D77}, 014019(2008)

\bibitem{penecostalba}
L. Cornalba, M.S. Costa, and J. Penedones,
{\it JHEP } {\bf 048}, 0806 (2008);

\bibitem{penedones}
J. Penedones,
{\it High Energy Scattering in the AdS/CFT Correspondence},
arXiv:0712.0802 [hep-th] 

\bibitem{lip86}
L.N. Lipatov,
{\it  Sov. Phys. JETP} {\bf 63}, 904 (1986). 

\bibitem{cornalba}
L. Cornalba,
{\it Eikonal methods in AdS/CFT: Regge theory and multi-reggeon exchange},
arXiv:0710.5480 [hep-th];


\bibitem{mobzor}
I. Balitsky, {\it ``High-Energy QCD and Wilson Lines''}, 
[hep-ph/0101042] 


\bibitem{npb96}
I. Balitsky, 
{\it Nucl. Phys.}  {\bf B463}, 99 (1996);
{\it ``Operator expansion for diffractive high-energy scattering''},
[hep-ph/9706411]; 
{\it Phys. Lett.} {\bf B518}, 235(2001).

\bibitem{yura}
Yu.V. Kovchegov,  
{\it Phys. Rev.} {\bf D60}, 034008 (1999);
{\it Phys. Rev.} {\bf D61},074018 (2000).

\bibitem{integral}
V.S. Fadin , M.I. Kotsky, and L.N. Lipatov,
{\it ``Gluon pair production in the quasimulti - Regge kinematics''},
[hep-ph/9704267]. 

\bibitem{fadinlo}
V.S. Fadin, R. Fiore,  {\it  Phys. Lett.} {\bf B661}, 139 (2008).

\bibitem{prd75}
I. Balitsky, {\it Phys. Rev.} {\bf D75}, 014001(2007),
Yu. V. Kovchegov and H. Weigert,
{\it Nucl. Phys.}  {\bf A784}, 188 (2007),


\bibitem{nlobfkl}
V.S. Fadin and L.N. Lipatov,
  {\it  Phys. Lett.} {\bf B429}, 127 (1998);
G. Camici and M. Ciafaloni,
  {\it Phys. Lett.} {\bf B430}, 349 (1998).
\end{thebibliography}
\end{document}